\documentclass[journal=jacsat,manuscript=article]{achemso}
\usepackage[version=3]{mhchem} 
\usepackage{color}

\setkeys{acs}{articletitle = true}

\author{Luke W. Bertels}
\affiliation{Department of Chemistry, University of California, Berkeley, California 94720, USA.}
\author{Joonho Lee}
\affiliation{Department of Chemistry, University of California, Berkeley, California 94720, USA.}
\author{Martin Head-Gordon}
\affiliation{Department of Chemistry, University of California, Berkeley, California 94720, USA.}
\affiliation{Chemical Sciences Division, Lawrence Berkeley National Laboratory, Berkeley, California 94720, USA.}
\email{mhg@cchem.berkeley.edu}

\title{Third-Order M{\o}ller-Plesset Perturbation Theory Made Useful? Choice of Orbitals and Scaling Greatly Improves Accuracy for Thermochemistry, Kinetics and Intermolecular Interactions}

\abbreviations{MP2, RI, OO, UHF, SOS, MP3, SCS, MP3, MP2.5, MP2.X, CCSD, CCSD(T), MR, aVTZ, RMSD, MSE, MIN, MAX}

\begin{document}

\begin{tocentry}





\centering
\includegraphics[scale=0.25]{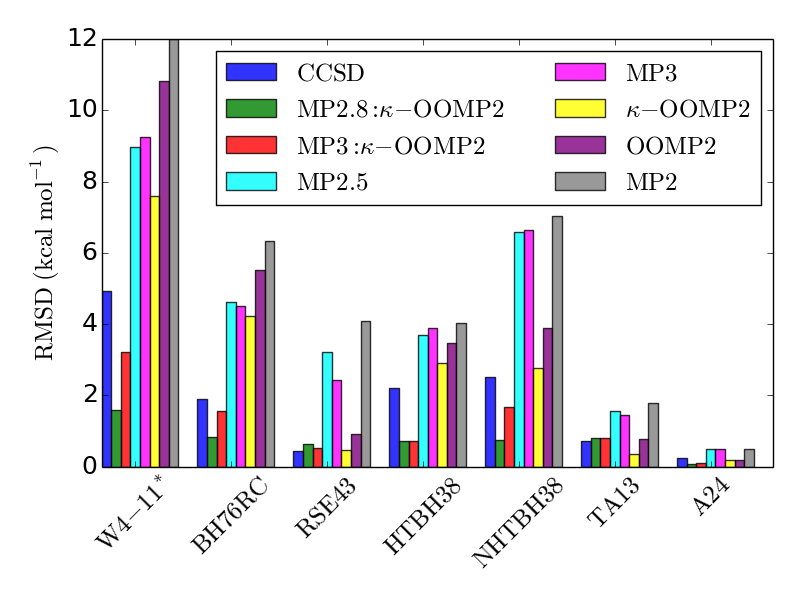}
\label{fig:rmsd_bar_plot}
\end{tocentry}

\begin{abstract}
 We develop and test methods that include second and third-order perturbation theory (MP3) using orbitals obtained from regularized orbital-optimized second-order perturbation theory, $\kappa$-OOMP2, denoted as MP3:$\kappa$-OOMP2. Testing MP3:$\kappa$-OOMP2 shows RMS errors that are 1.7 to 5 times smaller than  MP3 across 7 data sets. To do still better, empirical training of the scaling factors for the second- and third-order correlation energies and the regularization parameter on one of those data sets led to an unregularized scaled ($c_2=1.0$; $c_3=0.8$) denoted as MP2.8:$\kappa$-OOMP2. MP2.8:$\kappa$-OOMP2 yields significant additional improvement over MP3:$\kappa$-OOMP2 in 4 of 6 test data sets on thermochemistry, kinetics, and noncovalent interactions. Remarkably, these two methods outperform coupled cluster with singles and doubles in 5 of the 7 data sets considered, at greatly reduced cost (no $\mathcal{O}(N^6)$ iterations).
\end{abstract}


Single-reference second-order M{\o}ller-Plesset perturbation theory (MP2) is among the most popular correlated wavefunction methods in electronic structure theory, in part due to its economical $\mathcal{O}(N^5)$ scaling, where $N$ is the basis set size. 
\begin{equation}
E_{MP2} = -\frac{1}{4}\sum_{ijab}\frac{|\langle ij||ab\rangle |^{2}}{\Delta_{ij}^{ab}}
\label{eq:mp2}
\end{equation}
Equation \eqref{eq:mp2} gives the correlation energy for MP2, where $i$, $j$, \ldots represent occupied molecular orbitals, $a$, $b$, \ldots represent virtual molecular orbitals, and 
$\Delta_{ij}^{ab} = \varepsilon_{a} + \varepsilon_{b} - \varepsilon_{i} - \varepsilon_{j}$
is the (non-negative) energy denominator. The resolution-of-identity (RI) technique applied to MP2 has allowed for a much more widespread use due to the reduction of the prefactor in the overall computational cost of the algorithm.\cite{rimp2-1,rimp2-2} 

Orbital-optimized MP2 (OOMP2) methods were developed to improve the performance of MP2 for energies and other properties.\cite{sosoomp2,scsoomp2,oomp2-ccd}. For systems where the unrestricted Hartree-Fock (UHF) reference exhibits spin-contamination (artifical spin-symmetry breaking), the use of these reference orbitals can lead to catastrophic performance of MP2.\cite{mp2fail-1,mp2fail-2,mp2fail-3,mp2fail-4} OOMP2 can also be thought of as a relatively inexpensive way to approximate Br{\"u}ckner orbitals.\cite{sosoomp2} Orbital optimization at the MP2 level often reduces the level of spin-contamination and improves energetics.\cite{sosoomp2,scsoomp2,oomp2.5-2} 

Despite the benefits of OOMP2 described above, there are several unsatisfying characteristics of the method that limit its applicability. Orbital optimization at the MP2 level can produce divergent energy contributions due to small energy denominators. This is often observed when stretching bonds and leads to significant underestimation of harmonic vibrational frequencies.\cite{dregoomp2} OOMP2 also often fails to continuously transition from spin-restricted (R) to spin-unrestricted (U) solutions even when the U solution is lower in energy.\cite{stability} A continuous R to U transition requires a Coulson-Fischer point where the lowest eigenvalue of the R to U stability Hessian becomes zero.\cite{cfpoint} Resolution of this issue is necessary to reach the proper dissociation limit for bond-breaking curves. 

Our group has attempted to remedy these issues of OOMP2 through use of regularization to prevent divergence of the energy due to small energy denominators. The first of these approaches was to shift the energy denominator by a constant factor, $\delta$, so that $\Delta_{ij}^{ab}\leftarrow \Delta_{ij}^{ab}+\delta$.\cite{dregoomp2}
This simple form was able to partially resolve the two issues with OOMP2 described above. The regularization parameter, $\delta$, both prevents the energy expression from diverging and damps the correlation energy contribution from MP2, leading the method to more closely resemble the continuous R to U transition seen in the HF reference. Unfortunately, in the case of scaled opposite spin OOMP2 (SOS-OOMP2), Razban et al.\cite{dregsosoomp2} found the values of $\delta$ that could restore Coulson-Fischer points were very large and consequently led to poor performance on problems that are normally well-treated by MP2.  

Recently, two of us\cite{skoomp2} developed two new classes of orbital energy dependent regularizers for OOMP2, of which the most promising is denoted as $\kappa$-OOMP2. In $\kappa$-OOMP2, the matrix elements associated with small denominators are damped such that:
\begin{eqnarray}
\label{eq:kappamp2}
E_{MP2}(\kappa) = -\frac{1}{4}\sum_{ijab}\frac{|\langle ij||ab\rangle |^{2}}{\Delta_{ij}^{ab}}\left(1-e^{-\kappa\left(\Delta_{ij}^{ab}\right)}\right)^{2}
\end{eqnarray}
Unlike the case of $\delta$-OOMP2, for $\kappa$-OOMP2 the unregularized energy expression is recovered for large energy denominators, and in the limit of small energy denominators, the correlation energy contributions are zero.
Regularization parameters of $\kappa \leq 1.5 E_{h}^{-1}$ were found to restore Coulson-Fischer points for hydrogen, ethane, ethene, and ethyne bond-breaking curves. 
$\kappa$ was trained on the W4-11 set to suggest a value for general application.\cite{w4-11} The result, $\kappa=1.45 E_{h}^{-1}$, proved robust to further testing on the RSE43\cite{rse43-1} and TA13\cite{ta13} sets, and defines $\kappa$-OOMP2 as a replacement for OOMP2. Complex restricted (cR) and complex general (cG) orbital extensions of $\kappa$-OOMP2 have also been used to interrogate singlet biradicaloids\cite{crkoomp2} and the nature of symmetry breaking in fullerenes\cite{cgkoomp2}, respectively. 

The success and ubiquity of MP2 and OOMP2 have led several research groups to develop modified second-order methods aimed at improving energetics. Notable examples are spin-component-scaled MP2 (SCS-MP2)\cite{scsmp2-1,sosmp2-1,scsmp2-2,sosmp2-2,scsmp2-3,sosmp2-3} and orbital optimized SCS-MP2 (SCS-OOMP2)\cite{sosoomp2,scsoomp2} methods, which weight correlation contributions coming from same-spin and opposite-spin pairs of electrons differently. These techniques have also been applied to the second-order correlation contribution in several double-hybrid density functionals.\cite{scsdh-1,scsdh-2,scsdh-3,scsdh-4,scsoodh} A subset of these methods, scaled-opposite-spin MP2 (SOS-MP2)\cite{sosmp2-1,sosmp2-2,sosmp2-3} and SOS-OOMP2\cite{sosoomp2}, are notable in that they can be implemented via an overall $\mathcal{O}(N^4)$ computational cost. Another example are the attenuated MP2 methods that partially cancel basis set superposition errors with errors in MP2 itself to yield improved intermolecular interaction energies in finite basis sets.\cite{Goldey2012,Goldey2013,Goldey2014,Goldey2015} 
Density-fitting and Cholesky-decomposed variants of OOMP2 have also been developed to improve the computational efficiency of the method.\cite{df-oomp2} 

Inclusion of higher-order terms in the perturbative expansion provides another approach to improve energetics from MP2.
\begin{equation}
\label{eq:mp3-2}
\begin{split}
E_{MP3}=  \frac{1}{8}&\sum_{ijabcd}\left(t^{ab}_{ij}\right)^{*}\langle ab||cd\rangle t^{cd}_{ij} \\
 + \frac{1}{8}&\sum_{ijklab}\left(t^{ab}_{ij}\right)^{*}\langle kl||ij\rangle t^{ab}_{kl} \\
 - &\sum_{ijkabc} \left(t^{ab}_{ij}\right)^{*}\langle kb||ic\rangle t^{ac}_{kj}
\end{split}
\end{equation}
Equation \eqref{eq:mp3-2} gives the third-order M{\o}ller-Plesset (MP3) contribution to the correlation energy in the spin-orbital basis. MP3 formally scales as $\mathcal{O}(N^{6})$ with basis set size, and describes the leading interaction of first-order pair-correlations, $t^{ab}_{ij}$, with each other. However, despite the higher compute cost, MP3 only modestly improves MP2 results (e.g. see data presented later).
In passing we note that it is possible to utilize separable density fitting techniques such as tensor hypercontraction to achieve quartic scaling ($\mathcal O (N^4)$) MP3 (and also MP2) energy evaluation.\cite{hohenstein2012tensor}

Grimme\cite{scsmp3} developed a spin-component scaled MP3 (SCS-MP3) method that improved ground state energies over SCS-MP2 for reaction energies, atomization energies, ionization energies, and stretched geometries. This method applied an overall third-order correlation energy scaling factor of 0.25 in addition to the scaling factors for same-spin and opposite-spin components. For weak noncovalent interactions, application of MP3 has failed to substantially improve binding energies as compared to MP2.\cite{scsmp2-3,scsmp3,hobza-1,hobza-2,hobza-3,hobza-4,hobza-5} Hobza and coworkers\cite{hobza-3,hobza-4,hobza-5} proposed scaling the third-order correlation energy to interpolate between the MP2 and MP3, leading to the development of MP2.5 and MP2.X, in order to improve binding energies for noncovalent interactions. Following these successes, Bozkaya and coworkers\cite{oomp3,oomp3-2,oomp2.5-1,oomp2.5-2,oomp2.5-3} developed OOMP3 and OOMP2.5 and evaluated the performance of these methods on thermochemistry, kinetics, and noncovalent interactions. OOMP2.5 was shown to outperform coupled cluster theory with single and double excitations (CCSD)\cite{ccsd-1,ccsd-2} on reaction energies and barrier heights \cite{oomp2.5-2} and perform comparably to coupled cluster with single, double, and perturbative triple excitations [CCSD(T)]\cite{ccsd(t)} for noncovalent interactions\cite{oomp2.5-1}. These are very promising results. Analytic gradients for OOMP3, OOMP2.5, and their density-fitting variants have also been introduced.\cite{oomp3-grad,df-oomp3-grad} 

Following the recent success of regularized OOMP2 in treating inherent problems in OOMP2, we decided to explore the use of $\kappa$-OOMP2 orbitals at the MP3 level. At the same time, we wanted to see if $\kappa$ regularization in MP3 could improve the overall energetics. Beginning from $\kappa$-OOMP2 would allow this method to avoid energy divergences caused by small energy denominators.\cite{skoomp2} 
In $\kappa$-OOMP2, damping of the two-electron integrals leads to the following expression for the $t$-amplitudes:
\begin{equation}
\label{eq:kappat}
t_{ij}^{ab}(\kappa) = -\frac{\langle ab||ij\rangle}{\Delta_{ij}^{ab}}\left(1-e^{-\kappa\Delta_{ij}^{ab}}\right).
\end{equation}
Inserting Equation \eqref{eq:kappat} into Equation \eqref{eq:mp3-2} we arrive at a regularized third-order correlation energy expression, $E_{MP3}(\kappa)$.
Our first candidate ansatz involved calculating the $\kappa$-OOMP2 energy and applying a scaled single-shot $E_{MP3}(\kappa)$ correction. 
\begin{equation}
\label{eq:regmp2.x}
E_{c}(\kappa,\kappa_{2},c_{3}) = E_{\kappa-OOMP2}(\kappa)+c_{3}E_{MP3:\kappa-OOMP2}(\kappa_{2})
\end{equation}

As a second, alternative form, we considered using $\kappa$-OOMP2 (with $\kappa = 1.45 E_{h}^{-1}$) as a method to generate molecular orbitals for use in correlated calculations containing second and third order energies which could then be independently regularized and/or scaled: 
\begin{equation}
\label{eq:mp2.xkoomp2}
E_{c}(\kappa,\kappa_{2},c_{2},c_{3}) = c_{2}E_{MP2:\kappa-OOMP2}(\kappa_{2}) + c_{3}E_{MP3:\kappa-OOMP2}(\kappa_{2})
\end{equation}
In this second case the non-Brillouin singles contribution,$-\sum_{ia}f_{ia}^{2}/\Delta_{i}^{a}$, is included at second-order as $\kappa$-OOMP2 does not obey the Brillouin theorem. For simplicity (and ease of implementation), we do not include a non-Brillouin singles contribution at third-order.

We trained both energy functionals on the non-multireference (non-MR) subset of the W4-11 thermochemistry data set.\cite{w4-11} We excluded the MR points in the set from the training data because the single reference methods we are investigating should not be able to describe MR systems adequately.  Both reference and training calculations were performed using the aug-cc-pVTZ (aVTZ) basis set\cite{augccpvtz-1,augccpvtz-2,augccpvtz-3} and the corresponding RI basis\cite{riaugccpvtz,riccpvqz} without the frozen core approximation. Reference data were computed using CCSD(T)\cite{ccsd(t)}. All calculations were performed in a development version of Q-Chem\cite{qchem}. 

\begin{figure}
\centering
\includegraphics[scale=0.5]{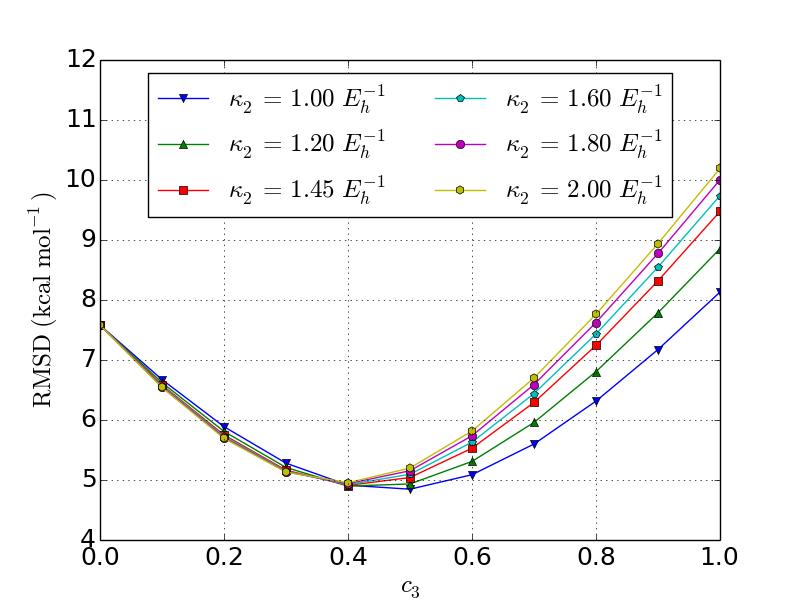}
\caption{Scans of the root mean square deviation on the non-MR subset of the W4-11 thermochemistry dataset, in kcal mol$^{-1}$, for the scaled, two regularization parameter correlation energy functional given in Equation \ref{eq:regmp2.x}. We fix $\kappa = 1.45 E_{h}^{-1}$. All calculations use the aVTZ basis; CCSD(T) is used for the reference values.
}
\label{fig:training1}
\end{figure}

Figure \ref{fig:training1} presents the root mean square deviations (RMSD) for scans of the $\kappa_{2}$ and $c_{3}$ parameters in the first model, as given by Equation \ref{eq:regmp2.x}. Overall, we see that stronger regularization at third-order (smaller $\kappa_{2}$) serves to lower the error on the training set. For $\kappa_{2} = 1.00 E_{h}^{-1}$, the optimal scaling parameter for the third-order regularized correlation energy is 0.5 for a RMSD of 4.85 kcal mol$^{-1}$. If instead one applies the same regularization parameter ($\kappa = 1.45 E_{h}^{-1}$) at second- and third-order, we find an optimal scaling parameter for the third order correlation energy of $c_{3} = 0.4$ with a RMSD of 4.91 kcal mol$^{-1}$. We note that $c_{3} = 0.0$ corresponds to $\kappa$-OOMP2 with a RMSD of 7.58 kcal mol$^{-1}$. Inclusion of scaled, regularized third-order correlation energy contributions reduces the RMSD of $\kappa$-OOMP2 by more than 2.5 kcal mol$^{-1}$, which is useful but not dramatic. 


\begin{figure}
    \centering
    \includegraphics[scale=0.5]{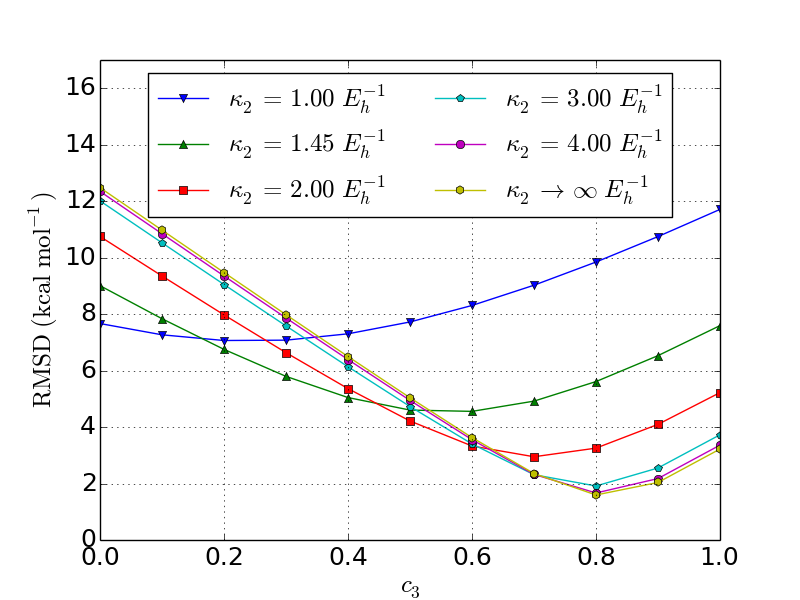}
    \caption{Scans of the root mean square deviation on the non-MR subset of the W4-11 thermochemical dataset, in kcal mol$^{-1}$, for the regularized, second- and third-order  correlation energy functional given in Equation \ref{eq:mp2.xkoomp2}. The optimal value of $c_{2}$ was found to be 1.0 for all $\kappa_{2}$ values plotted. SCF references were generated via $\kappa$-OOMP2 orbital optimization. The basis set used was aVTZ. Reference values are calculated at the CCSD(T)/aVTZ level of theory. }
    \label{fig:training2}
\end{figure}

Turning to the second form we considered, Figure \ref{fig:training2} presents the RMSDs for scans of the $\kappa_{2}$ and $c_{3}$ parameters in Equation \ref{eq:mp2.xkoomp2} with $c_{2} = 1.0$ (the optimal value for all $\kappa_{2}$ values plotted). For this ansatz, we see that the error relative to CCSD(T)/aVTZ is driven down quite dramatically by weakening the regularization (increasing $\kappa$). Indeed, perhaps surprisingly, we find that computing energies with unregularized MP2 and scaled unregularized MP3 provides the lowest error. In this case, the optimal $c_{3}$ parameter was found to be 0.8 and yields a RMSD of only 1.59 kcal mol$^{-1}$, which is close to chemical accuracy. We also observe that increasing the regularization strength decreases the optimal fraction of regularized third-order correlation energy. Impressed with the performance on this set, we chose this method, which we denote as MP2.8:$\kappa$-OOMP2, as our candidate for further evaluation. 

In order to assess the transferability of MP2.8:$\kappa$-OOMP2, we tested its performance on a series of benchmarks sets meant to encompass a variety of main group bonded and non-bonded interactions: the non-MR subset of the W4-11 set\cite{w4-11} (the training set), the BH76RC set\cite{bh76,htbh38,nhtbh38}, the RSE43 set\cite{rse43-1,rse43-2}, the HTBH38 set\cite{htbh38}, the NHTBH38 set\cite{nhtbh38}, the TA13 set\cite{ta13}, and the A24 set\cite{a24}. We compare the performance of MP2.8:$\kappa$-OOMP2 against an unscaled version of the method (MP3:$\kappa$-OOMP2), CCSD\cite{ccsd-1,ccsd-2}, MP2.5\cite{hobza-3}, MP3, $\kappa$-OOMP2\cite{skoomp2}, OOMP2, and MP2. Details of the computations (aVTZ basis, CCSDT(T) reference, no frozen core) are the same as given previously.

Table \ref{tab:w4-11} presents the the RMSDs, mean signed deviations (MSD), minimum deviations (MIN), and maximum deviations (MAX), in kcal mol$^{-1}$, for the non-MR subset of the W4-11 set (the training set). This set includes atomization energies (TAE140), bond dissociation energies (BDE99), heavy atom transfer energies (HAT707), nucleophilic substitution reaction energies (SN13), and isomerization energies (ISOMERIZATION20).\cite{w4-11} We see that CCSD has a RMSD of 4.94 kcal mol$^{-1}$ and a MSD of 1.49 kcal mol$^{-1}$. MP2, MP2.5 and MP3 on top of UHF orbitals yields RMSDs of 11.99, 8.97, and 9.24 kcal mol$^{-1}$, respectively. The use of $\kappa$-OOMP2 optimized orbitals for the computation of the MP3 energy reduces the error over the use of UHF orbitals by a remarkably large factor of 3. MP2.8/$\kappa$-OOMP2 yields a RMSD of 1.59 kcal mol$^{-1}$ and a MSD of -0.45 kcal mol$^{-1}$, which is 6 times smaller than MP3. This is also an improvement on $\kappa$-OOMP2 by a factor of 4. Also remarkably, both third-order methods computed using $\kappa$-OOMP2 optimized orbitals outperform CCSD on this data set. 

\begin{table}
\centering
\begin{tabular}{ccccc}
\hline
Method & RMSD & MSD & MIN & MAX \\
\hline
CCSD & 4.94 & 1.49 & -8.60 & 20.34 \\
MP2.8:$\kappa$-OOMP2 & 1.59 & -0.45 & -5.83 & 5.24 \\
MP3:$\kappa$-OOMP2 & 3.22 & 0.33 & -8.12 & 14.63 \\
MP2.5 & 8.97 & -2.85 & -40.24 & 24.87 \\
MP3 & 9.24 & -0.80 & -38.57 & 31.66 \\
$\kappa$-OOMP2 & 7.58 & -3.14 & -38.94 & 13.56 \\
OOMP2 & 10.82 & -3.50 & -48.81 & 17.87 \\
MP2 & 11.99 & -4.90 & -51.14 & 27.28 \\
\hline
\end{tabular}
\caption{Root mean square deviation, mean signed deviation, minimum deviation, and maximum deviation, in kcal mol$^{-1}$ for the non-MR subset of the W4-11 set. All calculations use the aVTZ basis; CCSD(T) is used for the reference values.}
\label{tab:w4-11}
\end{table}

 To validate the performance for thermochemistry outside of the training set, we tested MP2.8:$\kappa$-OOMP2 on the BH76RC\cite{bh76,htbh38,nhtbh38} and RSE43\cite{rse43-1,rse43-2} sets. The BH76RC set contains reaction energies for 30 reactions involved in the HTBH38 and NHTBH38 sets.\cite{bh76,htbh38,nhtbh38} On this set MP2.8:$\kappa$-OOMP2 outperforms all other methods surveyed with an RMSD of 0.84 kcal mol$^{-1}$ and a MSD of -0.14 kcal mol$^{-1}$. MP3:$\kappa$-OOMP2 also performs very well. Of the $\mathcal{O}(N^{5})$ methods, $\kappa$-OOMP2 provides the lowest RMSD while MP2 provides the lowest overall MSD. The largest absolute error using the canonical MP methods is for the \ce{H + F2 -> HF + H} reaction energy. This can be traced back to spin-contamination at the UHF level in the case of \ce{F2} with a $\langle S^2\rangle$ of 0.293. Both $\kappa$-OOMP2 and MP2.8:$\kappa$-OOMP2 show significant improvement on this case with errors of 1.425 kcal mol$^{-1}$ and 0.536 kcal mol$^{-1}$, respectively. MP2.8:$\kappa$-OOMP2 improves upon $\kappa$-OOMP2 in all but two cases in this set. 

\begin{table}
\centering
\begin{tabular}{ccccc}
\hline
Method & RMSD & MSD & MIN & MAX \\
\hline
CCSD & 1.905 & -0.645 & -7.175 & 1.977 \\
MP2.8:$\kappa$-OOMP2 & 0.835 & -0.143 & -1.465 & 1.534 \\
MP3:$\kappa$-OOMP2 & 1.574 & -0.437 & -6.213 & 1.228 \\
MP2.5 & 4.625 & -0.536 & -21.407 & 9.904 \\
MP3 & 4.511 & -0.975 & -21.604 & 4.232 \\
$\kappa$-OOMP2 & 4.220 & -0.276 & -9.763 & 11.856 \\
OOMP2 & 5.524 & 0.836 & -10.010 & 20.496 \\
MP2 & 6.341 & -0.098 & -21.211 & 15.577 \\
\hline
\end{tabular}
\caption{Root mean square deviation, mean signed deviation, minimum deviation, and maximum deviation, in kcal mol$^{-1}$, for the BH76RC set. }
\label{tab:bh76rc}
\end{table}

Table \ref{tab:rse43} contains benchmark results for the RSE43 set. The RSE43 set contains reaction energies for hydrogen abstraction from hydrocarbons by a methyl radical.\cite{rse43-1,rse43-2} For this set we see MP2.8:$\kappa$-OOMP2, with an RMSD of 0.63 kcal mol$^{-1}$ and a MSD of -0.54 kcal mol$^{-1}$, performs slightly worse than CCSD and MP3:$\kappa$-OOMP2. However the RMSD is still almost 4 times smaller than MP3. 
MP2.8:$\kappa$-OOMP2, MP3:$\kappa$-OOMP2, and OOMP2 all underestimate the reaction energies on average. For the $\mathcal{O}(N^5)$ methods, $\kappa$-OOMP2 outperforms OOMP2 and MP2 both in terms of RMSD and MSD. Several species in the set exhibit spin contamination, leading to failures of the canonical MP methods. 
\begin{table}
\centering
\begin{tabular}{ccccc}
\hline
Method & RMSD & MSD & MIN & MAX \\
\hline
CCSD & 0.446 & 0.316 & -0.815 & 0.973 \\
MP2.8:$\kappa$-OOMP2 & 0.634 & -0.538 & -1.726 & 0.050 \\
MP3:$\kappa$-OOMP2 & 0.521 & -0.416 & -1.550 & -0.002\\
MP2.5 & 3.234 & 1.907 & 0.061 & 12.899 \\
MP3 & 2.433 & 1.563 & 0.109 & 9.361 \\
$\kappa$-OOMP2 & 0.476 & 0.119 & -0.964 & 1.020 \\
OOMP2 & 0.922 & -0.607 & -2.261 & 0.478 \\
MP2 & 4.099 & 2.252 & -0.028 & 16.445 \\
\hline
\end{tabular}
\caption{Root mean square deviation, mean signed deviation, minimum deviation, and maximum deviation, in kcal mol$^{-1}$, for the RSE43 set. }
\label{tab:rse43}
\end{table}

To evaluate the performance of MP2.8/$\kappa$-OOMP2 on kinetics, we tested it on the HTBH38\cite{htbh38} and NHTBH38\cite{nhtbh38} data sets. The HTBH38 set contains forward and reverse barrier heights for 19 hydrogen transfer reactions.\cite{htbh38} Results for this set are presented in Table \ref{tab:htbh38}. On this set MP2.8:$\kappa$-OOMP2 (and MP3:$\kappa$-OOMP2) outperform the other methods surveyed with a RMSD of 0.71 kcal mol$^{-1}$ (and 0.73 kcal mol$^{-1}$), corresponding to chemical accuracy. The RMSDs are around 3 times smaller than that for CCSD. 
MP2, MP2.5, and MP3 overestimate the barrier heights in nearly all cases in the test set, with worst performances for the \ce{HF + H -> H2 + F}, \ce{HF + H -> H2 + F}, and \ce{OH + NH3 -> H2O + NH2} forward barriers, respectively. MP2.8:$\kappa$-OOMP2 improves significantly on these cases with barrier height errors of 0.24 kcal mol$^{-1}$ and -0.51 kcal mol$^{-1}$, respectively.

\begin{table}
\centering
\begin{tabular}{ccccc}
\hline
Method & RMSD & MSD & MIN & MAX \\
\hline
CCSD & 2.206 & 1.877 & -0.782 & 4.146 \\
MP2.8:$\kappa$-OOMP2 & 0.711 & -0.120 & -1.424 & 1.301 \\
MP3:$\kappa$-OOMP2 & 0.730 & 0.346 & -1.411 & 1.755 \\
MP2.5 & 3.686 & 3.246 & -0.273 & 7.323 \\
MP3 & 3.883 & 3.506 & 0.695 & 7.214 \\
$\kappa$-OOMP2 & 2.918 & 1.658 & -1.434 & 9.558 \\
OOMP2 & 3.479 & -0.952 & -7.152 & 8.566 \\
MP2 & 4.044 & 2.986 & -1.487 & 12.142 \\
\hline
\end{tabular}
\caption{Root mean square deviation, mean signed deviation, minimum deviation, and maximum deviation, in kcal mol$^{-1}$, for the HTBH38 set. }
\label{tab:htbh38}
\end{table}

Assessment data for the NHTBH38\cite{nhtbh38} set are presented in Table \ref{tab:nhtbh38}. The NHTBH38 set contains forward and reverse barrier heights for 19 non-hydrogen transfer reactions. On this set, MP2.8:$\kappa$-OOMP2 outperforms all other methods surveyed (RMSD of 0.76 kcal mol$^{-1}$), with MP3:$\kappa$-OOMP2 performing second best. The reduction in RMSD relative to MP3 is more than a factor of 8 for MP2.8:$\kappa$-OOMP2. Remarkably, both methods improve substantially upon CCSD, with the improvement being more than a factor of 3 for MP2.8:$\kappa$-OOMP2  
MP2, MP2.5, and MP3 all exhibit large errors in the barrier heights for the reactions \ce{H + N2O -> OH + N2}, \ce{H + F2 -> HF + F}, and \ce{CH3 + ClF -> CH3F + Cl}. For \ce{H + N2O -> OH + N2} and \ce{CH3 + ClF -> CH3F + Cl}, both forward and reverse barriers are overestimated due to spin-contamination of the UHF reference for the transition state structures. The UHF reference $\langle S^{2}\rangle$ values of 1.011 and 1.026, respectively, are corrected to  mean-field $\langle S^{2}\rangle$ values of 0.765 and 0.775, respectively, via the $\kappa$-OOMP2 orbital optimization procedure. For \ce{H + F2 -> HF + F}, the reverse barriers are overestimated by more than 20 kcal mol$^{-1}$ with MP2, MP2.5, and MP3 while errors in the forward barriers are of similar magnitude to other systems in the data set. Significant spin-contamination is present in the UHF reference for both \ce{F2} and the transition state structure, leading to a cancellation of errors in the forward barrier that is not seen in the reverse barrier. Orbital optimization with $\kappa$-OOMP2 helps to mitigate this spin-contamination, reducing the mean-field $\langle S^{2}\rangle$ values of 0.293 and 1.212, respectively, to 0.000 and 0.767, respectively. For all three of these reactions MP2.8:$\kappa$-OOMP2 gives errors that are reduced by a factor of 5-10 relative to MP2, MP2.5, and MP3. 

\begin{table}
\centering
\begin{tabular}{ccccc}
\hline
Method & RMSD & MSD & MIN & MAX \\
\hline
CCSD & 2.534 & 2.067 & 0.132 & 7.646 \\
MP2.8:$\kappa$-OOMP2 & 0.758 & 0.268 & -0.949 & 1.579 \\
MP3:$\kappa$-OOMP2 & 1.668 & 1.076 & -0.718 & 7.175 \\
MP2.5 & 6.590 & 4.763 & -0.328 & 24.455 \\
MP3 & 6.651 & 5.158 & 1.099 & 23.283 \\
$\kappa$-OOMP2 & 2.766 & 1.553 & -7.610 & 5.222 \\
OOMP2 & 3.901 & -1.650 & -18.495 & 2.315 \\
MP2 & 7.035 & 4.368 & -2.676 & 25.627 \\
\hline
\end{tabular}
\caption{Root mean square deviation, mean signed deviation, minimum deviation, and maximum deviation, in kcal mol$^{-1}$, for the NHTBH38 set.}
\label{tab:nhtbh38}
\end{table}

To round out the test suite we assessed the performance of MP2.8:$\kappa$-OOMP2 on two noncovalent interaction sets: the TA13 and A24 sets. The TA13 set contains 13 nonbonded interaction energies for radical closed-shell complexes.\cite{ta13} We apply a counterpoise correction to these interaction energies to mitigate basis set superposition error (BSSE). Assessment data for the TA13 set is presented in Table \ref{tab:ta13}. On this test set we see MP2.8:$\kappa$-OOMP2 performs almost as well as CCSD. MP2.8:$\kappa$-OOMP2 overbinds each interaction in the set, especially the \ce{H2O\hyphen Al} interaction which is overbound by 2.05 kcal mol$^{-1}$. Remarkably, $\kappa$-OOMP2 outperforms all methods surveyed on this set, with an RMSD of 0.35 kcal mol$^{-1}$. Table \ref{tab:a24} presents the counterpoise-corrected results for the A24 set. The A24 set contains noncovalent interaction energies for 24 closed-shell small molecule complexes.\cite{a24} MP2.8:$\kappa$-OOMP2 outperforms all other methods with a RMSD of 0.08 kcal mol$^{-1}$ and a MSD of 0.01 kcal mol$^{-1}$. For MP2, MP2.5, and MP3, artifactual spin-contamination at the UHF level causes underbinding for the ethene-ethene and ethene-ethyne dimers. For the ethene dimer, the MP2, MP2.5, and MP3 errors are in excess of 2 kcal mol$^{-1}$ while MP2.8:$\kappa$-OOMP2 reduces this error to 0.04 kcal mol$^{-1}$.

\begin{table}
\centering
\begin{tabular}{ccccc}
\hline
Method & RMSD & MSD & MIN & MAX \\
\hline
CCSD & 0.722 & 0.539 & -0.259 & 1.470 \\
MP2.8:$\kappa$-OOMP2 & 0.823 & -0.459 & -2.054 & -0.011 \\
MP3:$\kappa$-OOMP2 & 0.808 & -0.442 & -2.463 & 0.086 \\
MP2.5 & 1.559 & 0.276 & -3.888 & 3.708 \\
MP3 & 1.435 & 0.391 & -2.612 & 3.997 \\
$\kappa$-OOMP2 & 0.350 & -0.019 & -0.589 & 0.650 \\
OOMP2 & 0.789 & -0.149 & -1.938 & 1.370 \\
MP2 & 1.791 & 0.160 & -5.164 & 3.419 \\
\hline
\end{tabular}
\caption{Root mean square deviation, mean signed deviation, minimum deviation, and maximum deviation, in kcal mol$^{-1}$, for the TA13 set. }
\label{tab:ta13}
\end{table}

\begin{table}
\centering
\begin{tabular}{ccccc}
\hline
Method & RMSD & MSD & MIN & MAX \\
\hline
CCSD & 0.247 & 0.226 & 0.093 & 0.429 \\
MP2.8:$\kappa$-OOMP2 & 0.075 & 0.007 & -0.169 & 0.233 \\
MP3:$\kappa$-OOMP2 & 0.106 & 0.043 & -0.113 & 0.373 \\
MP2.5 & 0.492 & 0.132 & -0.113 & 2.303 \\
MP3 & 0.488 & 0.187 & -0.010 & 2.203 \\
$\kappa$-OOMP2 & 0.184 & -0.045 & -0.631 & 0.199 \\
OOMP2 & 0.193 & -0.131 & -0.475 & 0.063 \\
MP2 & 0.515 & 0.078 & -0.441 & 2.403 \\
\hline
\end{tabular}
\caption{Root mean square deviation, mean signed deviation, minimum deviation, and maximum deviation, in kcal mol$^{-1}$, for the A24 set. }
\label{tab:a24}
\end{table}

Considering all the data presented, let us summarize the main conclusions obtained from this work.
\begin{enumerate}
    \item At the MP2 level the choice of orbitals matters considerably, as is well known. In our work, for all 7 data sets considered, orbital optimized MP2 (OO-MP2) yields lower RMSD than MP2 (using unrestricted orbitals when necessary). Regularization via $\kappa=1.45$ has formal benefits in restoring Coulson-Fisher points. It also has practical benefits: $\kappa$-OOMP2 yields lower RMSD than OO-MP2 for all 7 data sets tested.
    \item Use of $\kappa$-OOMP2 orbitals improves MP3 results to a surprising  extent. MP3:$\kappa$-OOMP2 has lower RMSD than MP3 by factors ranging from 1.7 to more than 5 across the 7 datasets reported here. MP3:$\kappa$-OOMP2 is thus a far more robust method than MP3 itself, due to the reduced spin-contamination in $\kappa$-OOMP2 orbitals relative to HF orbitals.
    \item Developing a semi-empirical variant based on scaling the MP2 and MP3 contributions yielded $c_2=1.0$ and $c_3=0.8$ based on the non-MR part of the W4-11 dataset (no regularization is preferred). In transferability tests, this MP2.8:$\kappa$-OOMP2 method improves over MP3:$\kappa$-OOMP2 in 4 of our 6 test sets, with the other two being very similar.
    \item Remarkably, the results obtained with MP3:$\kappa$-OOMP2 and MP2.8:$\kappa$-OOMP2 produce lower RMSD than CCSD itself in 5 of the 7 datasets (the remaining two show no large failures). This indicates a level of performance that is beyond the physical content of MP3 theory, and involves some rather systematic cancellation of the effects due to connected triples.
    \item These improved MP3 methods are single reference of course, and the datasets considered here are suitable for single reference methods. Much poorer performance must be expected for systems where strong correlations are in play (perhaps with the exception of biradicaloids\cite{crkoomp2})
\end{enumerate}



The improved performance granted by the use of $\kappa$-OOMP2 optimized orbitals suggests future developments in electronic structure theory. It will be interesting to assess results across additional data sets, and explore the use of larger basis sets. We intend to explore scaled fourth-order perturbation approaches (MP4) and coupled cluster methods with $\kappa$-OOMP2 reference orbitals. The latter would be especially interesting in the context of nonvariational failures of CCSD(T). In a different direction, perhaps MP3 should be considered as an independent descriptor of electron correlation in double hybrid density functional theory, where MP2 is at present most widely used.\cite{scsdh-1,scsdh-2,scsdh-3,scsdh-4,scsoodh}
With the advances in integral compression techniques such as tensor hypercontraction, both MP2 and MP3 energy evaluations scale
quartically with system size.\cite{hohenstein2012tensor}
Incorporating this into the development of new double hybrid density functionals will
be a promising future direction.

\begin{acknowledgement}
This work was supported by the U.S. Department of Energy, Office of Basic Energy Science, and Office of Advanced Scientific Computing Research through the SciDAC program.
\end{acknowledgement}

\begin{suppinfo}
Supporting Information Available: raw energies for data sets

\end{suppinfo}

\bibliography{kappamp2.x.bib}

\providecommand{\latin}[1]{#1}
\providecommand*\mcitethebibliography{\thebibliography}
\csname @ifundefined\endcsname{endmcitethebibliography}
  {\let\endmcitethebibliography\endthebibliography}{}
\begin{mcitethebibliography}{64}
\providecommand*\natexlab[1]{#1}
\providecommand*\mciteSetBstSublistMode[1]{}
\providecommand*\mciteSetBstMaxWidthForm[2]{}
\providecommand*\mciteBstWouldAddEndPuncttrue
  {\def\EndOfBibitem{\unskip.}}
\providecommand*\mciteBstWouldAddEndPunctfalse
  {\let\EndOfBibitem\relax}
\providecommand*\mciteSetBstMidEndSepPunct[3]{}
\providecommand*\mciteSetBstSublistLabelBeginEnd[3]{}
\providecommand*\EndOfBibitem{}
\mciteSetBstSublistMode{f}
\mciteSetBstMaxWidthForm{subitem}{(\alph{mcitesubitemcount})}
\mciteSetBstSublistLabelBeginEnd
  {\mcitemaxwidthsubitemform\space}
  {\relax}
  {\relax}

\bibitem[Feyereisen \latin{et~al.}(1993)Feyereisen, Fitzgerald, and
  Komornicki]{rimp2-1}
Feyereisen,~M.; Fitzgerald,~G.; Komornicki,~A. Use of approximate integrals in
  ab initio theory. An application in MP2 energy calculations. \emph{Chem.
  Phys. Lett.} \textbf{1993}, \emph{208}, 359 -- 363\relax
\mciteBstWouldAddEndPuncttrue
\mciteSetBstMidEndSepPunct{\mcitedefaultmidpunct}
{\mcitedefaultendpunct}{\mcitedefaultseppunct}\relax
\EndOfBibitem
\bibitem[Bernholdt and Harrison(1996)Bernholdt, and Harrison]{rimp2-2}
Bernholdt,~D.~E.; Harrison,~R.~J. Large-scale correlated electronic structure
  calculations: the RI-MP2 method on parallel computers. \emph{Chem. Phys.
  Lett.} \textbf{1996}, \emph{250}, 477 -- 484\relax
\mciteBstWouldAddEndPuncttrue
\mciteSetBstMidEndSepPunct{\mcitedefaultmidpunct}
{\mcitedefaultendpunct}{\mcitedefaultseppunct}\relax
\EndOfBibitem
\bibitem[Lochan and Head-Gordon(2007)Lochan, and Head-Gordon]{sosoomp2}
Lochan,~R.~C.; Head-Gordon,~M. Orbital-optimized opposite-spin scaled
  second-order correlation: An economical method to improve the description of
  open-shell molecules. \emph{J. Chem. Phys.} \textbf{2007}, \emph{126},
  164101\relax
\mciteBstWouldAddEndPuncttrue
\mciteSetBstMidEndSepPunct{\mcitedefaultmidpunct}
{\mcitedefaultendpunct}{\mcitedefaultseppunct}\relax
\EndOfBibitem
\bibitem[Neese \latin{et~al.}(2009)Neese, Schwabe, Kossmann, Schirmer, and
  Grimme]{scsoomp2}
Neese,~F.; Schwabe,~T.; Kossmann,~S.; Schirmer,~B.; Grimme,~S. Assessment of
  Orbital-Optimized, Spin-Component Scaled Second-Order Many-Body Perturbation
  Theory for Thermochemistry and Kinetics. \emph{J. Chem. Theory Comput.}
  \textbf{2009}, \emph{5}, 3060--3073\relax
\mciteBstWouldAddEndPuncttrue
\mciteSetBstMidEndSepPunct{\mcitedefaultmidpunct}
{\mcitedefaultendpunct}{\mcitedefaultseppunct}\relax
\EndOfBibitem
\bibitem[Bozkaya \latin{et~al.}(2011)Bozkaya, Turney, Yamaguchi, Schaefer, and
  Sherrill]{oomp2-ccd}
Bozkaya,~U.; Turney,~J.~M.; Yamaguchi,~Y.; Schaefer,~H.~F.; Sherrill,~C.~D.
  Quadratically convergent algorithm for orbital optimization in the
  orbital-optimized coupled-cluster doubles method and in orbital-optimized
  second-order Møller-Plesset perturbation theory. \emph{J. Chem. Phys.}
  \textbf{2011}, \emph{135}, 104103\relax
\mciteBstWouldAddEndPuncttrue
\mciteSetBstMidEndSepPunct{\mcitedefaultmidpunct}
{\mcitedefaultendpunct}{\mcitedefaultseppunct}\relax
\EndOfBibitem
\bibitem[Farnell \latin{et~al.}(1983)Farnell, Pople, and Radom]{mp2fail-1}
Farnell,~L.; Pople,~J.~A.; Radom,~L. Structural predictions for open-shell
  systems: a comparative assessment of ab initio procedures. \emph{J. Phys.
  Chem.} \textbf{1983}, \emph{87}, 79--82\relax
\mciteBstWouldAddEndPuncttrue
\mciteSetBstMidEndSepPunct{\mcitedefaultmidpunct}
{\mcitedefaultendpunct}{\mcitedefaultseppunct}\relax
\EndOfBibitem
\bibitem[Nobes \latin{et~al.}(1987)Nobes, Pople, Radom, Handy, and
  Knowles]{mp2fail-2}
Nobes,~R.~H.; Pople,~J.~A.; Radom,~L.; Handy,~N.~C.; Knowles,~P.~J. Slow
  convergence of the M{\o}ller-Plesset perturbation series: the dissociation
  energy of hydrogen cyanide and the electron affinity of the cyano radical.
  \emph{Chem. Phys. Lett.} \textbf{1987}, \emph{138}, 481 -- 485\relax
\mciteBstWouldAddEndPuncttrue
\mciteSetBstMidEndSepPunct{\mcitedefaultmidpunct}
{\mcitedefaultendpunct}{\mcitedefaultseppunct}\relax
\EndOfBibitem
\bibitem[Gill \latin{et~al.}(1988)Gill, Pople, Radom, and Nobes]{mp2fail-3}
Gill,~P. M.~W.; Pople,~J.~A.; Radom,~L.; Nobes,~R.~H. Why does unrestricted
  M{\o}ller–Plesset perturbation theory converge so slowly for
  spin‐contaminated wave functions? \emph{J. Chem. Phys.} \textbf{1988},
  \emph{89}, 7307--7314\relax
\mciteBstWouldAddEndPuncttrue
\mciteSetBstMidEndSepPunct{\mcitedefaultmidpunct}
{\mcitedefaultendpunct}{\mcitedefaultseppunct}\relax
\EndOfBibitem
\bibitem[Jensen(1990)]{mp2fail-4}
Jensen,~F. A remarkable large effect of spin contamination on calculated
  vibrational frequencies. \emph{Chem. Phys. Lett.} \textbf{1990}, \emph{169},
  519 -- 528\relax
\mciteBstWouldAddEndPuncttrue
\mciteSetBstMidEndSepPunct{\mcitedefaultmidpunct}
{\mcitedefaultendpunct}{\mcitedefaultseppunct}\relax
\EndOfBibitem
\bibitem[Soyda{\c s} and Bozkaya(2015)Soyda{\c s}, and Bozkaya]{oomp2.5-2}
Soyda{\c s},~E.; Bozkaya,~U. Assessment of Orbital-Optimized MP2.5 for
  Thermochemistry and Kinetics: Dramatic Failures of Standard Perturbation
  Theory Approaches for Aromatic Bond Dissociation Energies and Barrier Heights
  of Radical Reactions. \emph{J. Chem. Theory Comput.} \textbf{2015},
  \emph{11}, 1564--1573\relax
\mciteBstWouldAddEndPuncttrue
\mciteSetBstMidEndSepPunct{\mcitedefaultmidpunct}
{\mcitedefaultendpunct}{\mcitedefaultseppunct}\relax
\EndOfBibitem
\bibitem[St{\" u}ck and Head-Gordon(2013)St{\" u}ck, and
  Head-Gordon]{dregoomp2}
St{\" u}ck,~D.; Head-Gordon,~M. Regularized orbital-optimized second-order
  perturbation theory. \emph{J. Chem. Phys.} \textbf{2013}, \emph{139},
  244109\relax
\mciteBstWouldAddEndPuncttrue
\mciteSetBstMidEndSepPunct{\mcitedefaultmidpunct}
{\mcitedefaultendpunct}{\mcitedefaultseppunct}\relax
\EndOfBibitem
\bibitem[Sharada \latin{et~al.}(2015)Sharada, St{\" u}ck, Sundstrom, Bell, and
  Head-Gordon]{stability}
Sharada,~S.~M.; St{\" u}ck,~D.; Sundstrom,~E.~J.; Bell,~A.~T.; Head-Gordon,~M.
  Wavefunction stability analysis without analytical electronic Hessians:
  application to orbital-optimised second-order M{\o}ller-Plesset theory and
  VV10-containing density functionals. \emph{Mol. Phys.} \textbf{2015},
  \emph{113}, 1802--1808\relax
\mciteBstWouldAddEndPuncttrue
\mciteSetBstMidEndSepPunct{\mcitedefaultmidpunct}
{\mcitedefaultendpunct}{\mcitedefaultseppunct}\relax
\EndOfBibitem
\bibitem[Coulson and Fischer(1949)Coulson, and Fischer]{cfpoint}
Coulson,~C.~A.; Fischer,~I. XXXIV. Notes on the molecular orbital treatment of
  the hydrogen molecule. \emph{Philos. Mag.} \textbf{1949}, \emph{40},
  386--393\relax
\mciteBstWouldAddEndPuncttrue
\mciteSetBstMidEndSepPunct{\mcitedefaultmidpunct}
{\mcitedefaultendpunct}{\mcitedefaultseppunct}\relax
\EndOfBibitem
\bibitem[Razban \latin{et~al.}(2017)Razban, St{\" u}ck, and
  Head-Gordon]{dregsosoomp2}
Razban,~R.~M.; St{\" u}ck,~D.; Head-Gordon,~M. Addressing first derivative
  discontinuities in orbital-optimised opposite-spin scaled second-order
  perturbation theory with regularisation. \emph{Mol. Phys.} \textbf{2017},
  \emph{115}, 2102--2109\relax
\mciteBstWouldAddEndPuncttrue
\mciteSetBstMidEndSepPunct{\mcitedefaultmidpunct}
{\mcitedefaultendpunct}{\mcitedefaultseppunct}\relax
\EndOfBibitem
\bibitem[Lee and Head-Gordon(2018)Lee, and Head-Gordon]{skoomp2}
Lee,~J.; Head-Gordon,~M. Regularized Orbital-Optimized Second-Order M{\"
  o}ller-Plesset Perturbation Theory: A Reliable Fifth-Order-Scaling Electron
  Correlation Model with Orbital Energy Dependent Regularizers. \emph{J. Chem.
  Theory Comput.} \textbf{2018}, \emph{14}, 5203--5219\relax
\mciteBstWouldAddEndPuncttrue
\mciteSetBstMidEndSepPunct{\mcitedefaultmidpunct}
{\mcitedefaultendpunct}{\mcitedefaultseppunct}\relax
\EndOfBibitem
\bibitem[Karton \latin{et~al.}(2011)Karton, Daon, and Martin]{w4-11}
Karton,~A.; Daon,~S.; Martin,~J.~M. W4-11: A high-confidence benchmark dataset
  for computational thermochemistry derived from first-principles W4 data.
  \emph{Chem. Phys. Lett.} \textbf{2011}, \emph{510}, 165 -- 178\relax
\mciteBstWouldAddEndPuncttrue
\mciteSetBstMidEndSepPunct{\mcitedefaultmidpunct}
{\mcitedefaultendpunct}{\mcitedefaultseppunct}\relax
\EndOfBibitem
\bibitem[Zipse(2006)]{rse43-1}
Zipse,~H. In \emph{Radicals in Synthesis I}; Gans{\"a}uer,~A., Ed.; Springer
  Berlin Heidelberg: Berlin, Heidelberg, 2006; pp 163--189\relax
\mciteBstWouldAddEndPuncttrue
\mciteSetBstMidEndSepPunct{\mcitedefaultmidpunct}
{\mcitedefaultendpunct}{\mcitedefaultseppunct}\relax
\EndOfBibitem
\bibitem[Tentscher and Arey(2013)Tentscher, and Arey]{ta13}
Tentscher,~P.~R.; Arey,~J.~S. Binding in Radical-Solvent Binary Complexes:
  Benchmark Energies and Performance of Approximate Methods. \emph{J. Chem.
  Theory Comput.} \textbf{2013}, \emph{9}, 1568--1579\relax
\mciteBstWouldAddEndPuncttrue
\mciteSetBstMidEndSepPunct{\mcitedefaultmidpunct}
{\mcitedefaultendpunct}{\mcitedefaultseppunct}\relax
\EndOfBibitem
\bibitem[Lee and Head-Gordon(2019)Lee, and Head-Gordon]{crkoomp2}
Lee,~J.; Head-Gordon,~M. Two single-reference approaches to singlet
  biradicaloid problems: Complex, restricted orbitals and approximate
  spin-projection combined with regularized orbital-optimized M{\o}ller-Plesset
  perturbation theory. \emph{J. Chem. Phys.} \textbf{2019}, \emph{150},
  244106\relax
\mciteBstWouldAddEndPuncttrue
\mciteSetBstMidEndSepPunct{\mcitedefaultmidpunct}
{\mcitedefaultendpunct}{\mcitedefaultseppunct}\relax
\EndOfBibitem
\bibitem[Lee and Head-Gordon(2019)Lee, and Head-Gordon]{cgkoomp2}
Lee,~J.; Head-Gordon,~M. Distinguishing artificial and essential symmetry
  breaking in a single determinant: approach and application to the \ce{C60},
  \ce{C36}, and \ce{C20} fullerenes. \emph{Phys. Chem. Chem. Phys.}
  \textbf{2019}, \emph{21}, 4763--4778\relax
\mciteBstWouldAddEndPuncttrue
\mciteSetBstMidEndSepPunct{\mcitedefaultmidpunct}
{\mcitedefaultendpunct}{\mcitedefaultseppunct}\relax
\EndOfBibitem
\bibitem[Grimme(2003)]{scsmp2-1}
Grimme,~S. Improved second-order M{\o}ller–Plesset perturbation theory by
  separate scaling of parallel- and antiparallel-spin pair correlation
  energies. \emph{J. Chem. Phys.} \textbf{2003}, \emph{118}, 9095--9102\relax
\mciteBstWouldAddEndPuncttrue
\mciteSetBstMidEndSepPunct{\mcitedefaultmidpunct}
{\mcitedefaultendpunct}{\mcitedefaultseppunct}\relax
\EndOfBibitem
\bibitem[Jung \latin{et~al.}(2004)Jung, Lochan, Dutoi, and
  Head-Gordon]{sosmp2-1}
Jung,~Y.; Lochan,~R.~C.; Dutoi,~A.~D.; Head-Gordon,~M. Scaled opposite-spin
  second order M{\o}ller–Plesset correlation energy: An economical electronic
  structure method. \emph{J. Chem. Phys.} \textbf{2004}, \emph{121},
  9793--9802\relax
\mciteBstWouldAddEndPuncttrue
\mciteSetBstMidEndSepPunct{\mcitedefaultmidpunct}
{\mcitedefaultendpunct}{\mcitedefaultseppunct}\relax
\EndOfBibitem
\bibitem[Grimme(2005)]{scsmp2-2}
Grimme,~S. Accurate Calculation of the Heats of Formation for Large Main Group
  Compounds with Spin-Component Scaled MP2 Methods. \emph{J. Phys. Chem. A}
  \textbf{2005}, \emph{109}, 3067--3077\relax
\mciteBstWouldAddEndPuncttrue
\mciteSetBstMidEndSepPunct{\mcitedefaultmidpunct}
{\mcitedefaultendpunct}{\mcitedefaultseppunct}\relax
\EndOfBibitem
\bibitem[Lochan \latin{et~al.}(2005)Lochan, Jung, and Head-Gordon]{sosmp2-2}
Lochan,~R.~C.; Jung,~Y.; Head-Gordon,~M. Scaled Opposite Spin Second Order
  M{\o}ller-Plesset Theory with Improved Physical Description of Long-Range
  Dispersion Interactions. \emph{J. Phys. Chem. A} \textbf{2005}, \emph{109},
  7598--7605\relax
\mciteBstWouldAddEndPuncttrue
\mciteSetBstMidEndSepPunct{\mcitedefaultmidpunct}
{\mcitedefaultendpunct}{\mcitedefaultseppunct}\relax
\EndOfBibitem
\bibitem[{Distasio Jr.} and Head-Gordon(2007){Distasio Jr.}, and
  Head-Gordon]{scsmp2-3}
{Distasio Jr.},~R.~A.; Head-Gordon,~M. Optimized spin-component scaled
  second-order M{\o}ller-Plesset perturbation theory for intermolecular
  interaction energies. \emph{Mol. Phys.} \textbf{2007}, \emph{105},
  1073--1083\relax
\mciteBstWouldAddEndPuncttrue
\mciteSetBstMidEndSepPunct{\mcitedefaultmidpunct}
{\mcitedefaultendpunct}{\mcitedefaultseppunct}\relax
\EndOfBibitem
\bibitem[Lochan \latin{et~al.}(2007)Lochan, Shao, and Head-Gordon]{sosmp2-3}
Lochan,~R.~C.; Shao,~Y.; Head-Gordon,~M. Quartic-Scaling Analytical Energy
  Gradient of Scaled Opposite-Spin Second-Order M{\o}ller-Plesset Perturbation
  Theory. \emph{J. Chem. Theory Comput.} \textbf{2007}, \emph{3},
  988--1003\relax
\mciteBstWouldAddEndPuncttrue
\mciteSetBstMidEndSepPunct{\mcitedefaultmidpunct}
{\mcitedefaultendpunct}{\mcitedefaultseppunct}\relax
\EndOfBibitem
\bibitem[Grimme(2006)]{scsdh-1}
Grimme,~S. Semiempirical hybrid density functional with perturbative
  second-order correlation. \emph{J. Chem. Phys.} \textbf{2006}, \emph{124},
  034108\relax
\mciteBstWouldAddEndPuncttrue
\mciteSetBstMidEndSepPunct{\mcitedefaultmidpunct}
{\mcitedefaultendpunct}{\mcitedefaultseppunct}\relax
\EndOfBibitem
\bibitem[Chai and Head-Gordon(2009)Chai, and Head-Gordon]{scsdh-2}
Chai,~J.-D.; Head-Gordon,~M. Long-range corrected double-hybrid density
  functionals. \emph{J. Chem. Phys.} \textbf{2009}, \emph{131}, 174105\relax
\mciteBstWouldAddEndPuncttrue
\mciteSetBstMidEndSepPunct{\mcitedefaultmidpunct}
{\mcitedefaultendpunct}{\mcitedefaultseppunct}\relax
\EndOfBibitem
\bibitem[Kozuch and Martin(2011)Kozuch, and Martin]{scsdh-3}
Kozuch,~S.; Martin,~J. M.~L. DSD-PBEP86: in search of the best double-hybrid
  DFT with spin-component scaled MP2 and dispersion corrections. \emph{Phys.
  Chem. Chem. Phys.} \textbf{2011}, \emph{13}, 20104--20107\relax
\mciteBstWouldAddEndPuncttrue
\mciteSetBstMidEndSepPunct{\mcitedefaultmidpunct}
{\mcitedefaultendpunct}{\mcitedefaultseppunct}\relax
\EndOfBibitem
\bibitem[Mardirossian and Head-Gordon(2018)Mardirossian, and
  Head-Gordon]{scsdh-4}
Mardirossian,~N.; Head-Gordon,~M. Survival of the most transferable at the top
  of Jacob’s ladder: Defining and testing the $\omega$B97M(2) double hybrid
  density functional. \emph{J. Chem. Phys.} \textbf{2018}, \emph{148},
  241736\relax
\mciteBstWouldAddEndPuncttrue
\mciteSetBstMidEndSepPunct{\mcitedefaultmidpunct}
{\mcitedefaultendpunct}{\mcitedefaultseppunct}\relax
\EndOfBibitem
\bibitem[Najibi and Goerigk(2018)Najibi, and Goerigk]{scsoodh}
Najibi,~A.; Goerigk,~L. A Comprehensive Assessment of the Effectiveness of
  Orbital Optimization in Double-Hybrid Density Functionals in the Treatment of
  Thermochemistry, Kinetics, and Noncovalent Interactions. \emph{J. Phys. Chem.
  A} \textbf{2018}, \emph{122}, 5610--5624\relax
\mciteBstWouldAddEndPuncttrue
\mciteSetBstMidEndSepPunct{\mcitedefaultmidpunct}
{\mcitedefaultendpunct}{\mcitedefaultseppunct}\relax
\EndOfBibitem
\bibitem[Goldey and Head-Gordon(2012)Goldey, and Head-Gordon]{Goldey2012}
Goldey,~M.; Head-Gordon,~M. Attenuating Away the Errors in Inter- and
  Intramolecular Interactions from Second-Order Moller-Plesset Calculations in
  the Small Aug-cc-pVDZ Basis Set. \emph{J. Phys. Chem. Lett.} \textbf{2012},
  \emph{3}, 3592--3598\relax
\mciteBstWouldAddEndPuncttrue
\mciteSetBstMidEndSepPunct{\mcitedefaultmidpunct}
{\mcitedefaultendpunct}{\mcitedefaultseppunct}\relax
\EndOfBibitem
\bibitem[Goldey \latin{et~al.}(2013)Goldey, Dutoi, and Head-Gordon]{Goldey2013}
Goldey,~M.; Dutoi,~A.; Head-Gordon,~M. Attenuated second-order Moller-Plesset
  perturbation theory: performance within the aug-cc-pVTZ basis. \emph{Phys.
  Chem. Chem. Phys.} \textbf{2013}, \emph{15}, 15869--15875\relax
\mciteBstWouldAddEndPuncttrue
\mciteSetBstMidEndSepPunct{\mcitedefaultmidpunct}
{\mcitedefaultendpunct}{\mcitedefaultseppunct}\relax
\EndOfBibitem
\bibitem[Goldey and Head-Gordon(2014)Goldey, and Head-Gordon]{Goldey2014}
Goldey,~M.; Head-Gordon,~M. Separate Electronic Attenuation Allowing a
  Spin-Component-Scaled Second-Order Moller-Plesset Theory to Be Effective for
  Both Thermocheniistry and Noncovalent Interactions. \emph{J. Phys. Chem. B}
  \textbf{2014}, \emph{118}, 6519--6525\relax
\mciteBstWouldAddEndPuncttrue
\mciteSetBstMidEndSepPunct{\mcitedefaultmidpunct}
{\mcitedefaultendpunct}{\mcitedefaultseppunct}\relax
\EndOfBibitem
\bibitem[Goldey \latin{et~al.}(2015)Goldey, Belzunces, and
  Head-Gordon]{Goldey2015}
Goldey,~M.~B.; Belzunces,~B.; Head-Gordon,~M. Attenuated MP2 with a Long-Range
  Dispersion Correction for Treating Nonbonded Interactions. \emph{J. Chem.
  Theor. Comput.} \textbf{2015}, \emph{11}, 4159--4168\relax
\mciteBstWouldAddEndPuncttrue
\mciteSetBstMidEndSepPunct{\mcitedefaultmidpunct}
{\mcitedefaultendpunct}{\mcitedefaultseppunct}\relax
\EndOfBibitem
\bibitem[Bozkaya(2014)]{df-oomp2}
Bozkaya,~U. Orbital-Optimized Second-Order Perturbation Theory with
  Density-Fitting and Cholesky Decomposition Approximations: An Efficient
  Implementation. \emph{J. Chem. Theory Comput.} \textbf{2014}, \emph{10},
  2371--2378\relax
\mciteBstWouldAddEndPuncttrue
\mciteSetBstMidEndSepPunct{\mcitedefaultmidpunct}
{\mcitedefaultendpunct}{\mcitedefaultseppunct}\relax
\EndOfBibitem
\bibitem[Hohenstein \latin{et~al.}(2012)Hohenstein, Parrish, and
  Mart{\'\i}nez]{hohenstein2012tensor}
Hohenstein,~E.~G.; Parrish,~R.~M.; Mart{\'\i}nez,~T.~J. Tensor hypercontraction
  density fitting. I. Quartic scaling second-and third-order M{\o}ller-Plesset
  perturbation theory. \emph{J. Chem. Phys.} \textbf{2012}, \emph{137},
  044103\relax
\mciteBstWouldAddEndPuncttrue
\mciteSetBstMidEndSepPunct{\mcitedefaultmidpunct}
{\mcitedefaultendpunct}{\mcitedefaultseppunct}\relax
\EndOfBibitem
\bibitem[Grimme(2003)]{scsmp3}
Grimme,~S. Improved third-order Møller–Plesset perturbation theory. \emph{J.
  Comput. Chem.} \textbf{2003}, \emph{24}, 1529--1537\relax
\mciteBstWouldAddEndPuncttrue
\mciteSetBstMidEndSepPunct{\mcitedefaultmidpunct}
{\mcitedefaultendpunct}{\mcitedefaultseppunct}\relax
\EndOfBibitem
\bibitem[Gr{\'a}fov{\'a} \latin{et~al.}(2010)Gr{\'a}fov{\'a}, Pito{\u n}{\'a}k,
  {\u R}ez{\'a}{\u c}, and Hobza]{hobza-1}
Gr{\'a}fov{\'a},~L.; Pito{\u n}{\'a}k,~M.; {\u R}ez{\'a}{\u c},~J.; Hobza,~P.
  Comparative Study of Selected Wave Function and Density Functional Methods
  for Noncovalent Interaction Energy Calculations Using the Extended S22 Data
  Set. \emph{J. Chem. Theory Comput.} \textbf{2010}, \emph{6}, 2365--2376\relax
\mciteBstWouldAddEndPuncttrue
\mciteSetBstMidEndSepPunct{\mcitedefaultmidpunct}
{\mcitedefaultendpunct}{\mcitedefaultseppunct}\relax
\EndOfBibitem
\bibitem[Riley \latin{et~al.}(2012)Riley, Platts, {\u R}ez{\'a}{\u c}, Hobza,
  and Hill]{hobza-2}
Riley,~K.~E.; Platts,~J.~A.; {\u R}ez{\'a}{\u c},~J.; Hobza,~P.; Hill,~J.~G.
  Assessment of the Performance of MP2 and MP2 Variants for the Treatment of
  Noncovalent Interactions. \emph{J. Phys. Chem. A} \textbf{2012}, \emph{116},
  4159--4169\relax
\mciteBstWouldAddEndPuncttrue
\mciteSetBstMidEndSepPunct{\mcitedefaultmidpunct}
{\mcitedefaultendpunct}{\mcitedefaultseppunct}\relax
\EndOfBibitem
\bibitem[Pito{\u n}{\'a}k \latin{et~al.}(2009)Pito{\u n}{\'a}k, Neogr{\'a}dy,
  {\u C}ern{\'y}, Grimme, and Hobza]{hobza-3}
Pito{\u n}{\'a}k,~M.; Neogr{\'a}dy,~P.; {\u C}ern{\'y},~J.; Grimme,~S.;
  Hobza,~P. Scaled MP3 Non-Covalent Interaction Energies Agree Closely with
  Accurate CCSD(T) Benchmark Data. \emph{ChemPhysChem} \textbf{2009},
  \emph{10}, 282--289\relax
\mciteBstWouldAddEndPuncttrue
\mciteSetBstMidEndSepPunct{\mcitedefaultmidpunct}
{\mcitedefaultendpunct}{\mcitedefaultseppunct}\relax
\EndOfBibitem
\bibitem[Riley \latin{et~al.}(2012)Riley, {\u R}ez{\'a}{\u c}, and
  Hobza]{hobza-4}
Riley,~K.~E.; {\u R}ez{\'a}{\u c},~J.; Hobza,~P. The performance of MP2.5 and
  MP2.X methods for nonequilibrium geometries of molecular complexes.
  \emph{Phys. Chem. Chem. Phys.} \textbf{2012}, \emph{14}, 13187--13193\relax
\mciteBstWouldAddEndPuncttrue
\mciteSetBstMidEndSepPunct{\mcitedefaultmidpunct}
{\mcitedefaultendpunct}{\mcitedefaultseppunct}\relax
\EndOfBibitem
\bibitem[Sedlak \latin{et~al.}(2013)Sedlak, Riley, {\u R}ez{\'a}{\u c}, Pito{\u
  n}{\'a}k, and Hobza]{hobza-5}
Sedlak,~R.; Riley,~K.~E.; {\u R}ez{\'a}{\u c},~J.; Pito{\u n}{\'a}k,~M.;
  Hobza,~P. MP2.5 and MP2.X: Approaching CCSD(T) Quality Description of
  Noncovalent Interaction at the Cost of a Single CCSD Iteration.
  \emph{ChemPhysChem} \textbf{2013}, \emph{14}, 698--707\relax
\mciteBstWouldAddEndPuncttrue
\mciteSetBstMidEndSepPunct{\mcitedefaultmidpunct}
{\mcitedefaultendpunct}{\mcitedefaultseppunct}\relax
\EndOfBibitem
\bibitem[Bozkaya(2011)]{oomp3}
Bozkaya,~U. Orbital-optimized third-order M{\o}ller-Plesset perturbation theory
  and its spin-component and spin-opposite scaled variants: Application to
  symmetry breaking problems. \emph{J. Chem. Phys.} \textbf{2011}, \emph{135},
  224103\relax
\mciteBstWouldAddEndPuncttrue
\mciteSetBstMidEndSepPunct{\mcitedefaultmidpunct}
{\mcitedefaultendpunct}{\mcitedefaultseppunct}\relax
\EndOfBibitem
\bibitem[Soyda{\c s} and Bozkaya(2013)Soyda{\c s}, and Bozkaya]{oomp3-2}
Soyda{\c s},~E.; Bozkaya,~U. Assessment of Orbital-Optimized Third-Order
  M{\o}ller-Plesset Perturbation Theory and Its Spin-Component and
  Spin-Opposite Scaled Variants for Thermochemistry and Kinetics. \emph{J.
  Chem. Theory Comput.} \textbf{2013}, \emph{9}, 1452--1460\relax
\mciteBstWouldAddEndPuncttrue
\mciteSetBstMidEndSepPunct{\mcitedefaultmidpunct}
{\mcitedefaultendpunct}{\mcitedefaultseppunct}\relax
\EndOfBibitem
\bibitem[Bozkaya and Sherrill(2014)Bozkaya, and Sherrill]{oomp2.5-1}
Bozkaya,~U.; Sherrill,~C.~D. Orbital-optimized MP2.5 and its analytic
  gradients: Approaching CCSD(T) quality for noncovalent interactions. \emph{J.
  Chem. Phys.} \textbf{2014}, \emph{141}, 204105\relax
\mciteBstWouldAddEndPuncttrue
\mciteSetBstMidEndSepPunct{\mcitedefaultmidpunct}
{\mcitedefaultendpunct}{\mcitedefaultseppunct}\relax
\EndOfBibitem
\bibitem[Bozkaya(2016)]{oomp2.5-3}
Bozkaya,~U. Orbital-Optimized MP3 and MP2.5 with Density-Fitting and Cholesky
  Decomposition Approximations. \emph{J. Chem. Theory Comput.} \textbf{2016},
  \emph{12}, 1179--1188\relax
\mciteBstWouldAddEndPuncttrue
\mciteSetBstMidEndSepPunct{\mcitedefaultmidpunct}
{\mcitedefaultendpunct}{\mcitedefaultseppunct}\relax
\EndOfBibitem
\bibitem[Purvis and Bartlett(1982)Purvis, and Bartlett]{ccsd-1}
Purvis,~G.~D.; Bartlett,~R.~J. A full coupled‐cluster singles and doubles
  model: The inclusion of disconnected triples. \emph{J. Chem. Phys.}
  \textbf{1982}, \emph{76}, 1910--1918\relax
\mciteBstWouldAddEndPuncttrue
\mciteSetBstMidEndSepPunct{\mcitedefaultmidpunct}
{\mcitedefaultendpunct}{\mcitedefaultseppunct}\relax
\EndOfBibitem
\bibitem[Scuseria \latin{et~al.}(1988)Scuseria, Janssen, and Schaefer]{ccsd-2}
Scuseria,~G.~E.; Janssen,~C.~L.; Schaefer,~H.~F. An efficient reformulation of
  the closed‐shell coupled cluster single and double excitation (CCSD)
  equations. \emph{J. Chem. Phys.} \textbf{1988}, \emph{89}, 7382--7387\relax
\mciteBstWouldAddEndPuncttrue
\mciteSetBstMidEndSepPunct{\mcitedefaultmidpunct}
{\mcitedefaultendpunct}{\mcitedefaultseppunct}\relax
\EndOfBibitem
\bibitem[Raghavachari \latin{et~al.}(1989)Raghavachari, Trucks, Pople, and
  Head-Gordon]{ccsd(t)}
Raghavachari,~K.; Trucks,~G.~W.; Pople,~J.~A.; Head-Gordon,~M. A fifth-order
  perturbation comparison of electron correlation theories. \emph{Chem. Phys.
  Lett.} \textbf{1989}, \emph{157}, 479 -- 483\relax
\mciteBstWouldAddEndPuncttrue
\mciteSetBstMidEndSepPunct{\mcitedefaultmidpunct}
{\mcitedefaultendpunct}{\mcitedefaultseppunct}\relax
\EndOfBibitem
\bibitem[Bozkaya(2013)]{oomp3-grad}
Bozkaya,~U. Analytic energy gradients for the orbital-optimized third-order
  M{\o}ller–Plesset perturbation theory. \emph{J. Chem. Phys.} \textbf{2013},
  \emph{139}, 104116\relax
\mciteBstWouldAddEndPuncttrue
\mciteSetBstMidEndSepPunct{\mcitedefaultmidpunct}
{\mcitedefaultendpunct}{\mcitedefaultseppunct}\relax
\EndOfBibitem
\bibitem[Bozkaya(2018)]{df-oomp3-grad}
Bozkaya,~U. Analytic energy gradients for orbital-optimized MP3 and MP2.5 with
  the density-fitting approximation: An efficient implementation. \emph{J.
  Comput. Chem.} \textbf{2018}, \emph{39}, 351--360\relax
\mciteBstWouldAddEndPuncttrue
\mciteSetBstMidEndSepPunct{\mcitedefaultmidpunct}
{\mcitedefaultendpunct}{\mcitedefaultseppunct}\relax
\EndOfBibitem
\bibitem[Dunning(1989)]{augccpvtz-1}
Dunning,~T.~H. Gaussian basis sets for use in correlated molecular
  calculations. I. The atoms boron through neon and hydrogen. \emph{J. Chem.
  Phys.} \textbf{1989}, \emph{90}, 1007--1023\relax
\mciteBstWouldAddEndPuncttrue
\mciteSetBstMidEndSepPunct{\mcitedefaultmidpunct}
{\mcitedefaultendpunct}{\mcitedefaultseppunct}\relax
\EndOfBibitem
\bibitem[Kendall \latin{et~al.}(1992)Kendall, Dunning, and
  Harrison]{augccpvtz-2}
Kendall,~R.~A.; Dunning,~T.~H.; Harrison,~R.~J. Electron affinities of the
  first‐row atoms revisited. Systematic basis sets and wave functions.
  \emph{J. Chem. Phys.} \textbf{1992}, \emph{96}, 6796--6806\relax
\mciteBstWouldAddEndPuncttrue
\mciteSetBstMidEndSepPunct{\mcitedefaultmidpunct}
{\mcitedefaultendpunct}{\mcitedefaultseppunct}\relax
\EndOfBibitem
\bibitem[Woon and Dunning(1993)Woon, and Dunning]{augccpvtz-3}
Woon,~D.~E.; Dunning,~T.~H. Gaussian basis sets for use in correlated molecular
  calculations. III. The atoms aluminum through argon. \emph{J. Chem. Phys.}
  \textbf{1993}, \emph{98}, 1358--1371\relax
\mciteBstWouldAddEndPuncttrue
\mciteSetBstMidEndSepPunct{\mcitedefaultmidpunct}
{\mcitedefaultendpunct}{\mcitedefaultseppunct}\relax
\EndOfBibitem
\bibitem[Weigend \latin{et~al.}(2002)Weigend, K{\"o}hn, and
  H{\"a}ttig]{riaugccpvtz}
Weigend,~F.; K{\"o}hn,~A.; H{\"a}ttig,~C. Efficient use of the correlation
  consistent basis sets in resolution of the identity MP2 calculations.
  \emph{J. Chem. Phys.} \textbf{2002}, \emph{116}, 3175--3183\relax
\mciteBstWouldAddEndPuncttrue
\mciteSetBstMidEndSepPunct{\mcitedefaultmidpunct}
{\mcitedefaultendpunct}{\mcitedefaultseppunct}\relax
\EndOfBibitem
\bibitem[H{\"a}ttig(2005)]{riccpvqz}
H{\"a}ttig,~C. Optimization of auxiliary basis sets for RI-MP2 and RI-CC2
  calculations: Core-valence and quintuple-$\zeta$ basis sets for H to Ar and
  QZVPP basis sets for Li to Kr. \emph{Phys. Chem. Chem. Phys.} \textbf{2005},
  \emph{7}, 59--66\relax
\mciteBstWouldAddEndPuncttrue
\mciteSetBstMidEndSepPunct{\mcitedefaultmidpunct}
{\mcitedefaultendpunct}{\mcitedefaultseppunct}\relax
\EndOfBibitem
\bibitem[Shao \latin{et~al.}(2015)Shao, Gan, Epifanovsky, Gilbert, Wormit,
  Kussmann, Lange, Behn, Deng, Feng, Ghosh, Goldey, Horn, Jacobson, Kaliman,
  Khaliullin, Ku{\'s}, Landau, Liu, Proynov, Rhee, Richard, Rohrdanz, Steele,
  Sundstrom, III, Zimmerman, Zuev, Albrecht, Alguire, Austin, Beran, Bernard,
  Berquist, Brandhorst, Bravaya, Brown, Casanova, Chang, Chen, Chien, Closser,
  Crittenden, Diedenhofen, Jr., Do, Dutoi, Edgar, Fatehi, Fusti-Molnar,
  Ghysels, Golubeva-Zadorozhnaya, Gomes, Hanson-Heine, Harbach, Hauser,
  Hohenstein, Holden, Jagau, Ji, Kaduk, Khistyaev, Kim, Kim, King, Klunzinger,
  Kosenkov, Kowalczyk, Krauter, Lao, Laurent, Lawler, Levchenko, Lin, Liu,
  Livshits, Lochan, Luenser, Manohar, Manzer, Mao, Mardirossian, Marenich,
  Maurer, Mayhall, Neuscamman, Oana, Olivares-Amaya, O'Neill, Parkhill,
  Perrine, Peverati, Prociuk, Rehn, Rosta, Russ, Sharada, Sharma, Small, Sodt,
  Stein, St{\"u}ck, Su, Thom, Tsuchimochi, Vanovschi, Vogt, Vydrov, Wang,
  Watson, Wenzel, White, Williams, Yang, Yeganeh, Yost, You, Zhang, Zhang,
  Zhao, Brooks, Chan, Chipman, Cramer, III, Gordon, Hehre, Klamt, III, Schmidt,
  Sherrill, Truhlar, Warshel, Xu, Aspuru-Guzik, Baer, Bell, Besley, Chai,
  Dreuw, Dunietz, Furlani, Gwaltney, Hsu, Jung, Kong, Lambrecht, Liang,
  Ochsenfeld, Rassolov, Slipchenko, Subotnik, Voorhis, Herbert, Krylov, Gill,
  and Head-Gordon]{qchem}
Shao,~Y. \latin{et~al.}  Advances in molecular quantum chemistry contained in
  the Q-Chem 4 program package. \emph{Mol. Phys.} \textbf{2015}, \emph{113},
  184--215\relax
\mciteBstWouldAddEndPuncttrue
\mciteSetBstMidEndSepPunct{\mcitedefaultmidpunct}
{\mcitedefaultendpunct}{\mcitedefaultseppunct}\relax
\EndOfBibitem
\bibitem[Goerigk and Grimme(2010)Goerigk, and Grimme]{bh76}
Goerigk,~L.; Grimme,~S. A General Database for Main Group Thermochemistry,
  Kinetics, and Noncovalent Interactions: Assessment of Common and
  Reparameterized (meta-)GGA Density Functionals. \emph{J. Chem. Theory
  Comput.} \textbf{2010}, \emph{6}, 107--126\relax
\mciteBstWouldAddEndPuncttrue
\mciteSetBstMidEndSepPunct{\mcitedefaultmidpunct}
{\mcitedefaultendpunct}{\mcitedefaultseppunct}\relax
\EndOfBibitem
\bibitem[Zhao \latin{et~al.}(2005)Zhao, Lynch, and Truhlar]{htbh38}
Zhao,~Y.; Lynch,~B.~J.; Truhlar,~D.~G. Multi-coefficient extrapolated density
  functional theory for thermochemistry and thermochemical kinetics.
  \emph{Phys. Chem. Chem. Phys.} \textbf{2005}, \emph{7}, 43--52\relax
\mciteBstWouldAddEndPuncttrue
\mciteSetBstMidEndSepPunct{\mcitedefaultmidpunct}
{\mcitedefaultendpunct}{\mcitedefaultseppunct}\relax
\EndOfBibitem
\bibitem[Zhao \latin{et~al.}(2005)Zhao, Gonz{\'a}lez-Garc{\'i}a, and
  Truhlar]{nhtbh38}
Zhao,~Y.; Gonz{\'a}lez-Garc{\'i}a,~N.; Truhlar,~D.~G. Benchmark Database of
  Barrier Heights for Heavy Atom Transfer, Nucleophilic Substitution,
  Association, and Unimolecular Reactions and Its Use to Test Theoretical
  Methods. \emph{J. Phys. Chem. A} \textbf{2005}, \emph{109}, 2012--2018\relax
\mciteBstWouldAddEndPuncttrue
\mciteSetBstMidEndSepPunct{\mcitedefaultmidpunct}
{\mcitedefaultendpunct}{\mcitedefaultseppunct}\relax
\EndOfBibitem
\bibitem[Goerigk \latin{et~al.}(2017)Goerigk, Hansen, Bauer, Ehrlich, Najibi,
  and Grimme]{rse43-2}
Goerigk,~L.; Hansen,~A.; Bauer,~C.; Ehrlich,~S.; Najibi,~A.; Grimme,~S. A look
  at the density functional theory zoo with the advanced GMTKN55 database for
  general main group thermochemistry, kinetics and noncovalent interactions.
  \emph{Phys. Chem. Chem. Phys.} \textbf{2017}, \emph{19}, 32184--32215\relax
\mciteBstWouldAddEndPuncttrue
\mciteSetBstMidEndSepPunct{\mcitedefaultmidpunct}
{\mcitedefaultendpunct}{\mcitedefaultseppunct}\relax
\EndOfBibitem
\bibitem[{\u R}ez{\' a}{\u c} and Hobza(2013){\u R}ez{\' a}{\u c}, and
  Hobza]{a24}
{\u R}ez{\' a}{\u c},~J.; Hobza,~P. Describing Noncovalent Interactions beyond
  the Common Approximations: How Accurate Is the "Gold Standard", CCSD(T) at
  the Complete Basis Set Limit? \emph{J. Chem. Theory Comput.} \textbf{2013},
  \emph{9}, 2151--2155\relax
\mciteBstWouldAddEndPuncttrue
\mciteSetBstMidEndSepPunct{\mcitedefaultmidpunct}
{\mcitedefaultendpunct}{\mcitedefaultseppunct}\relax
\EndOfBibitem
\end{mcitethebibliography}

\end{document}